\documentclass[prl,preprint]{revtex4}
\usepackage{amssymb}

\input{tcilatex}

\begin{document}

\title{Quasiparticle relaxation dynamics in heavy fermion compounds}
\author{J. Demsar$^{\dagger }$, R.D. Averitt$^{\dagger }$, K.H. Ahn$%
^{\dagger }$, M.J. Graf$^{\dagger }$, S.A. Trugman$^{\dagger }$, V.V. Kabanov%
$^{\ddagger }$, J.L. Sarrao$^{\dagger }$, A.J. Taylor$^{\dagger }$}
\affiliation{$^{\dagger }$ Los Alamos National Laboratory, Mail Stop K764, Los Alamos,
New Mexico 87545\\
$^{\ddagger }$ \textquotedblleft J. Stefan\textquotedblright\ Institute,
Jamova 39, SI-1000, Ljubljana, Slovenia}

\begin{abstract}
We present the first femtosecond studies of electron-phonon (e-ph)
thermalization in heavy fermion compounds. The e-ph thermalization time $%
\tau _{ep}$ increases below the Kondo temperature by more than two orders of
magnitude as T = 0 K is approached. Analysis using the two-temperature model
and numerical simulations based on Boltzmann's equations suggest that this
anomalous slowing down of the e-ph thermalization derives from the large
electronic specific heat and the suppresion of scattering between heavy
electrons and phonons.
\end{abstract}

\maketitle

Recent experiments have demonstrated that femtosecond time-resolved optical
spectroscopy is a sensitive tool to probe the low energy electronic
structure of strongly correlated electron systems[1-4], complementing
conventional time-averaged frequency-domain methods. In these experiments, a
femtosecond laser pulse excites a non-thermal electron distribution. This
non-thermal distribution rapidly thermalizes through electron-electron (e-e)
interactions resulting in a change in the occupied density of states (DOS)
in proximity to the Fermi energy (E$_{f}$). Therefore, by measuring
photoinduced reflectivity or transmissivity dynamics as a function of
temperature (T), it is possible to sensitively probe the nature of the
electronic ground state. For example, femtosecond measurements of the
carrier relaxation dynamics of high-T$_{c}$ superconductors and charge
density wave compounds have provided new insights into the low energy
electronic structure of these materials [1-3]. What is particularly
important is that even though the probe photon wavelength in these
experiments ranges from the far-infrared\cite{Averitt}, the near-IR\cite%
{Kabanov}, up to several eV\cite{Stevens}, the relaxation dynamics on
identical samples is the same\cite{Averitt}, supporting the basic idea\cite%
{Kabanov} that dynamic photoinduced reflectivity measurements, in many
instances, probe relaxation and recombination processes of quasiparticles in
the vicinity of E$_{f}$.

In this Letter, we present the first studies of carrier relaxation dynamics
in heavy fermion (HF) compounds using femtosecond time-resolved optical
spectroscopy, aiming to elucidate the effect of localized $f$-electrons\cite%
{Reviews} on the quasiparticle relaxation dynamics. We have measured the
time-resolved photoinduced reflectivity $\Delta $R/R dynamics as a function
of temperature on the series of HF compounds Yb\textit{X}Cu$_{4}$ (\textit{X}%
=Ag, Cd, In) \cite{Sarrao} in comparison to their non-magnetic counterparts
LuXCu$_{4}$. Our results reveal that the carrier relaxation dynamics are
extremely sensitive to the low energy DOS near E$_{F}$. In particular, in HF
compounds the relaxation time $\tau _{r}$ shows\ a hundred-fold increase
between the Kondo temperature (T$_{K}$) and $10$ K, while in the
non-magnetic analogues $\tau _{r}$ is nearly constant, similar to
conventional metals like Ag, Au, and Cu\cite{Groenevald}. Our analysis shows
that the relaxation dynamics can be attributed to e-ph thermalization, and
that the anomalous slowing down of the e-ph thermalization stems from the
large electronic specific heat in HF compounds and suppression of e-ph
scattering within the peak in the enhanced density of states near E$_{f}$.

In the following, we focus on YbAgCu$_{4}$ (a prototypical HF system with $%
T_{K}\sim 100$ K and low temperature Sommerfeld coefficient $\gamma \sim 210$
mJ/mol K$^{2}$)\cite{Sarrao} compared to its non-magnetic counterpart LuAgCu$%
_{4}$ ($\gamma \sim 10$ mJ/mol K$^{2}$). The experiments were performed on
freshly polished flux-grown single crystals\cite{Sarrao}. We used a standard
pump-probe set-up with a mode-locked Ti:Sapphire laser producing 20 fs
pulses centered at 800 nm (photon energy $\hbar \omega _{ph}\simeq 1.5$ eV)
with an 80 MHz repetition rate. The photoinduced (PI) changes in
reflectivity $\Delta $R/R were measured using a photodiode and lock-in
detection. The pump fluence was kept below 0.1 $\mu $J/cm$^{2}$ to minimize
the overall heating of the illuminated spot\cite{Mihailovic}, while the
pump/probe intensity ratio was $\sim 30$. Steady-state heating effects were
accounted for as described in \cite{Mihailovic}, yielding an uncertainty in
temperature of $\pm $3 K (in all the data the temperature increase of the
illuminated spot has been accounted for).

Figure 1 presents the PI reflectivity traces on the two compounds at several
temperatures between $\approx 10$ K and 300 K. The relaxation dynamics of
the non-HF compound LuAgCu$_{4}$ display a very weak temperature dependence,
with $\Delta $R/R recovering on a sub-picosecond timescale at all T. The
dynamics are similar to regular metals such as Au and Ag \cite{Groenevald},
where the recovery is predominantly due to e-ph thermalization. In contrast,
Fig. 1(b) shows that for YbAgCu, the quasiparticle dynamics are strongly $T$%
-dependent. Specifically, above $\sim 140$ K, the recovery time $\tau _{r}$
(determined by a exp$(-t/\tau _{r})$ fit to the data) is virtually $T$%
-independent but increases by more than two orders of magnitude as T $%
\rightarrow $ 0 K. We have measured similar dynamics on YbCdCu$_{4}$ ($%
T_{K}\sim 100$ K, $\gamma \sim 200$ mJ/mol K%
${{}^2}$%
). Furthermore, a similar divergence of $\tau _{r}$ occurs for CeCoIn$_{5}$
below $\approx 60$ K, implying that the observed increase in the relaxation
time starting at $\sim T_{K}$ and its subsequent divergence as T $%
\rightarrow $ 0 K is a generic feature of HF compounds and derives from
their low energy electronic structure.

The rise-time dynamics are also different in the two compounds. For LuAgCu$%
_{4}$, the rise-time is $\sim 100$ fs at all temperatures. This is again
similar to what has been measured on conventional metals and reflects the
time it takes for the initially created high energy quasiparticles to
thermalize towards E$_{f}$. Above $\sim 25$ K, YbAgCu$_{4}$ displays a
similar (fast) rise-time. Below this temperature the rise-time increases
and, as the semi-log plot in Fig. 1(b) reveals, becomes two-exponential at
the lowest temperatures. Similar behavior also occurs for CeCoIn$_{5}$, but
is absent in YbCdCu$_{4}$ indicating a strong dependence on the details of
the low energy electronic structure in HF compounds. While noting the
presence of these anomalous rise-time dynamics, further systematic studies
are needed to obtain a more complete understanding.\cite{note2} In the
following, we focus on the anomalous temperature dependence of recovery
dynamics below T$_{K}$ which seem to be a general feature of HF compounds.

In conventional metals, the initial photoinduced change in the reflectivity
arises from changes in occupation near $E_{f}$ after e-e thermalization.
Subsequently, the PI reflectivity recovery dynamics proceed on a picosecond
timescale governed by e-ph thermalization\cite{Groenevald}. The two
temperature model (TTM) serves as a useful starting point in describing e-ph
thermalization in metals.\cite{Anisimov,Kaganov,allen}. The TTM describes
the time evolution of the electron ($T_{e}$) and lattice ($T_{l}$)\
temperatures by two coupled differential equations\cite{Anisimov,Groenevald}%
. In the low photoexcitation energy density limit, as in our case, when $%
T_{e}-T_{l}\ll T_{l}$ over the entire temperature range, the set of two
coupled differential equations can be linearized resulting in the following
expression for the e-ph thermalization time%
\begin{equation}
\tau _{ep}^{-1}=g(C_{e}^{-1}+C_{l}^{-1})\text{ \ .}  \label{Eq4}
\end{equation}%
Here $C_{e}$ and $C_{l}$ are the electronic and lattice specific heats,
respectively, and $g(T_{l})$ is the e-ph coupling function\cite%
{Anisimov,Groenevald}. In\ the case of simple metals, when the electron
bandwidth is much larger than the Debye energy $\hbar \omega
_{D}=k_{B}\Theta _{D}$, and using the Debye model for the e-ph interaction, $%
g(T)$ has particularly simple form. It is given by $g(T)=dG(T)/dT$, where 
\cite{Kaganov,Groenevald} 
\begin{equation}
G(T)=4g_{\infty }\left( \frac{T}{\Theta _{D}}\right)
^{5}\dint\nolimits_{0}^{\Theta _{D}/T}dx\frac{x^{4}}{e^{x}-1}\ \chi (x,T).
\label{GTsimp}
\end{equation}%
Here $g_{\infty }$ is termed the e-ph coupling constant, while $\chi (x,T)$
is included to account for the variation in the electronic DOS, $D_{e}\left(
\epsilon \right) $, and the normalized e-ph scattering strength $F(\epsilon
,\epsilon ^{\prime })$, over the energy range $E_{f}\pm \hbar \omega _{D}$.
It can be shown using Fermi's golden rule that%
\begin{equation}
\chi (x,T)=\frac{1}{\xi }\int\limits_{-\infty }^{\infty }d\epsilon \frac{%
D_{e}\left( \epsilon \right) D_{e}\left( \epsilon ^{\prime }\right)
F(\epsilon ,\epsilon ^{\prime })}{D_{0}^{2}}\left\{ f_{0}\left( \epsilon
\right) -f_{0}\left( \epsilon ^{\prime }\right) \right\} ,  \label{ChiFunc}
\end{equation}%
where $\epsilon ^{\prime }=\epsilon +\xi $, and $\xi =xT$ and $f_{0}$ is the
Fermi-Dirac distribution. In metals like Au or Ag, $D_{el}\left( \epsilon
\right) $ and $F$ are approximately constant in this energy range, i.e. $%
D_{e}\left( \epsilon \right) =D_{0}$ and $F=1$, giving $\chi \equiv 1$. $%
g_{\infty }$ is typically $10^{15}-10^{16}$ W/mol K (e.g. for Cu $g_{\infty
}=6.2\times 10^{15}$ W/mol K corresponding to $g(300K)=1\times 10^{17}$ W/m$%
^{3}$K \cite{Elsayed}).

At high temperatures ($T>\Theta _{D}$), $\tau _{ep}\left( T\right) $ given
by the TTM has been found to describe the temperature as well as
photoexcitation intensity dependence of measured $\tau _{r}\left( T\right) $ 
\cite{Groenevald,Elsayed}. Moreover, since the absolute value of $\tau _{ep}$
is determined by \emph{a single parameter} $g_{\infty }$, the technique has
been successfully used to determine the dimensionless e-ph coupling
constants $\lambda $ in superconductors\cite{allen,Brorson}. However, at low
temperatures ($T\lesssim \Theta _{D}/5$) the TTM prediction of $\tau _{ep}$ $%
\propto T^{-3}$ has never been observed in metals\cite{Groenevald} -
instead, at low temperatures $\tau _{r}$ saturates at a constant value. The
discrepancy between the experimental results and the TTM was found to be due
to the fact that the TTM neglects e-e thermalization processes (by
implicitly assuming that a Fermi-Dirac distribution is created instantly
following photoexcitation). From simulations using coupled Boltzmann
equations, Groeneveld \textit{et al.}\cite{Groenevald} showed that this
discrepancy is due to the fact that at low temperatures the e-e and e-ph
thermalization times are comparable. Since $\tau _{ep}\propto T$ above $%
\approx \Theta _{D}/6$\cite{allen,Groenevald}, while $\tau _{ee}\varpropto
T^{-2}$ - see Eq.(16) of Ref.\cite{Groenevald}, the TTM is expected to fail
at low temperatures where $\tau _{ee}\gtrapprox \tau _{ep}$.

In Fig.2 we plot the T-dependence of $\tau _{r}$ on LuAgCu$_{4}$ (solid
circles), together with the TTM prediction for $\tau _{ep}$ (dashed line)
given by Eqs.(\ref{Eq4},\ref{GTsimp}) with $\Theta _{D}=280$ K\cite{Sarrao},
measured $C_{e}(T)$ and $C_{l}(T)$ - see inset to Fig. 2, and $g_{\infty
}=2.6\times 10^{15}$ W/mol K. Similar to Au or Ag\cite{Groenevald}, we find
good agreement at $T\gtrsim $ 200 K, while below 40 K instead of showing a $%
\tau _{ep}\propto T^{-3}$ divergence, $\tau _{r}$ saturates.

In order to explain the discrepancy, we have carried out numerical
simulations using coupled Boltzmann equations\cite{Groenevald,fullana}.
Here, for example, the net phonon absorption by electrons with energy $%
\epsilon $ is represented by $\left[ \frac{df_{\epsilon }}{dt}\right]
_{ep}^{abs}=\int d\omega K_{ep}S(\epsilon ,\omega )D_{p}\left( \omega
\right) D_{e}\left( \epsilon +\omega \right) ,$ where $D_{e}(\epsilon )$ and 
$D_{p}(\omega )$ are the electron and phonon DOS,\ and $S(\epsilon ,\omega
)=f_{\epsilon +\omega }(1-f_{\epsilon })-b_{\omega }(f_{\epsilon
}-f_{\epsilon +\omega })$, with $f$ and $b$ being the electron and phonon
distribution functions, respectively. $K_{ep}$ in the above equation and $%
K_{ee}$ in e-e scattering represent the square of the scattering matrix
element, combined with all other numerical factors\cite{Groenevald,fullana}.
When performing numerical simulations, a thermal phonon distribution ($%
b_{t=0}=b_{0}(T_{l})$) and a non-thermal electron distribution ($%
f_{t=0}=f_{0}(T_{e})\pm \delta f$) was taken as the initial condition just
after the laser pulse\cite{Groenevald}, while $\tau _{ep}$ is found by
fitting the total electron energy versus time curve to an exponential decay
function. The initial perturbation $\delta f$ is around $10^{-5}\sim 10^{-3}$
for the energy range between 0.10 $\sim $ 0.15 eV above and below $E_{F}$,
which is small enough that the increase in the temperature after e-ph
thermalization is less than 1 K over the whole temperature range -
consistent with the small excitation intensity used in the experiment\cite%
{numerics1}. The phonon and electron DOS used in the simulation were chosen
such that they fit the specific heat data (i.e. for the phonon DOS, we use
the Debye model $D_{p}(\omega )\sim \omega ^{2}$ with $\hbar \omega _{D}=$24
meV, while $D_{e}(E_{F})\approx 2.1$ eV$^{-1}$ f.u.$^{-1}$spin$^{-1}$). The
result of the simulation using the absolute value of $K_{ep}$ $=0.93$ ps$%
^{-1}$eV and $K_{ee}/K_{ep}=700$ is plotted by open circles\cite{numerics2}
in Fig.2. As expected, the simulation gives the same result as the TTM at
high-T, while at low-T $\tau _{ep}$ saturates in agreement with the
experimental $\tau _{r}$.

Figure 3 shows $\tau _{r}(T)$ obtained on YbAgCu$_{4}$. At high temperatures
($T>T_{K}$) the value of\ $\tau _{r}$ is similar to LuAgCu$_{4}$. At low
temperatures, however, $\tau _{r}$ increases by more than 2 orders of
magnitude. Since heavy fermions are characterized by a peak in the DOS at $%
E_{f}$\cite{Reviews}, the appropriate $D_{el}(\epsilon )$ should be used
when modeling $\tau _{ep}(T)$. In our calculation we used $D_{e}(\epsilon
)=D_{peak}\exp [-(\epsilon /\Delta )^{2}]+D_{0}$, where $D_{peak}=70$ eV$%
^{-1}$f.u.$^{-1}$spin$^{-1}$, $\Delta =13$ meV and $D_{0}=2.1$ eV$^{-1}$f.u.$%
^{-1}$spin$^{-1}$ (identical to the value for LuAgCu$_{4}$). It reproduces
the experimental T-dependence of $C_{e}$, as shown in the inset to Fig.3.
For simplicity we choose $E_{f}$ at the center of the peak, so that the
chemical potential is constant. Since $D_{e}(E_{f})$ is almost two orders of
magnitude larger than in LuAgCu$_{4}$ we expect that the e-e thermalization
is much faster in YbAgCu$_{4}$, and that the TTM would be valid at the
lowest temperatures.

The calculated $\tau _{ep}(T)$ using Eq.(\ref{GTsimp}) is plotted in Fig.3
by the dashed line. Here the approximate $C_{e}\left( T\right) $ and $%
C_{l}\left( T\right) $ were used, $g_{\infty }$ was taken to be the same as
for LuAgCu$_{4}$, while $\chi \left( x,T\right) $ was evaluated explicitly
for the above $D_{e}(\epsilon )$ and $F=1$. Since $\tau _{ep}^{-1}$ $%
\varpropto $ $D_{e}$, and $D_{e}(E_{f})\gg D_{0}$ the result is not
surprising, implying that the simple TTM cannot account for the observed
dramatic increase in $\tau _{r}$ at low temperatures. We should note that
neither the value of the e-ph coupling constant $g_{\infty }$ nor $D_{0}$,
which determine the absolute value of $\tau _{ep}$, are necessarily the same
in Yb\textit{X}Cu$_{4}$ and Lu\textit{X}Cu$_{4}$. However, even if the e-ph
coupling is 10 times smaller in YbAgCu$_{4}$ compared to LuAgCu$_{4}$ (which
would give 10 times larger value of $\tau _{ep}$ - as plotted by dashed line
in Figure 3), the observed T-dependence of $\tau _{r}$ still cannot be
accounted for.

In order to account for the observed $\tau _{r}\left( T\right) $ we have to
consider the nature of the electronic states within the peak in the DOS. In
heavy fermions the peak in $D_{e}(\epsilon )$ originates from hybridization
of the localized $f$-levels with the conduction band electrons\cite{Reviews}%
. We hypothesize that the e-ph scattering within the DOS peak is suppressed,
since the band dispersion near $E_{f}$ is much weaker than in regular
metals. Therefore it is quite possible that the Fermi velocity $v_{F}$ is
smaller than the sound velocity $v_{s}$, in which case \textit{momentum and
energy conservation} prohibit e-ph scattering when both initial and final
electron states lie within the energy range where $v_{F}<v_{s}$. Assuming a
parabolic band with $E_{F}\sim T_{K}\sim $ 100 K, and 0.85 carriers per
formula unit\cite{Hall}, one obtains $v_{F}\sim $ 4 km/sec, while the
longitudinal sound velocity for YbIn$_{1-x}$Ag$_{x}$Cu$_{4}$ ($x<0.3$) is $%
\approx $ 4.4 km/sec along [111] direction\cite{sound} (similar $v_{s}$ is
expected for YbAgCu$_{4}$). Even though a parabolic dispersion relation is
just a rough approximation, and a direct measurement such as de Haas-van
Alphen effect is required to obtain $v_{F}$, our simple estimate supports
this idea.

Using this hypothesis, good agreement with the data can be obtained. $\tau
_{ep}(T)$ obtained by numerical simulations based on Boltzmann equations
with $K_{ep}$ set to $0$ for processes where the initial and final
electronic state are in the range of $-24<\epsilon <24$ meV (i.e. within the
DOS peak), and $K_{ep}=0.23$ ps$^{-1}$eV otherwise, is plotted by open
circles in Fig. 3. Even better agreement with the data is obtained from the
TTM, assuming that the e-ph interaction strength $F\left( \epsilon ,\epsilon
^{\prime }\right) $ entering Eq.(\ref{ChiFunc}) smoothly vanishes as $%
\epsilon $ and $\epsilon ^{\prime }\rightarrow E_{f}$, accounting for $v_{F}$
variation (and thus $v_{F}<v_{s}$ condition) across the Fermi surface. This
is implemented into the TTM simulation by approximating the factor $%
D_{e}\left( \epsilon \right) D_{e}\left( \epsilon ^{\prime }\right) F\left(
\epsilon ,\epsilon ^{\prime }\right) $ in Eq.(\ref{ChiFunc}) with the
symmetrized function $(D_{e}\left( \epsilon \right) D_{ie}\left( \epsilon
^{\prime }\right) +$ $D_{e}\left( \epsilon ^{\prime }\right) D_{ie}\left(
\epsilon \right) -D_{ie}\left( \epsilon \right) D_{ie}\left( \epsilon
^{\prime }\right) )$, where $D_{ie}\left( \epsilon \right) =D_{0}-D_{0}\exp
[-(\epsilon /\Delta ^{\prime })^{2}]$. The resulting $\tau _{ep}(T)$, using $%
\Delta ^{\prime }=24$ meV, and $g_{\infty }=4.5\times 10^{14}$ W/mol K is
plotted by the solid line in Fig. 3. Indeed, extremely good agreement with
the data is obtained, even though $\tau _{r}$ spans more than two orders of
magnitude\cite{note3}.

With the hypothesis that e-ph scattering is suppressed in the DOS peak, the
experimental observation of anomalous T-dependence of $\tau _{r}$ can be
understood. Namely, at $T<T_{K}$ $C_{e}\left( T\right) $ increases
dramatically compared to normal metals. On the other hand e-ph relaxation
becomes more and more difficult as temperature is lowered since most of the
electron relaxation should occur within the DOS peak, where the e-ph
scattering is blocked by \textit{energy and momentum} conservation.
Therefore, the thermalization between electrons and the lattice occurs very
slowly, giving rise to the divergent $\tau _{ep}$ below $T_{K}$. While the
presented model explains the main features of the data, i.e. the low-T
divergence of $\tau _{ep}$, there are still several issues requiring further
experimental and theoretical effort.

For example, in the simulations we considered a temperature independent peak
in the DOS, assuming that many-body and correlation effects can be described
by effective, T-independent model parameters. This may be an
oversimplification of the physics of heavy-fermion systems. However, the
relaxation time simulations and specific heat calculations of our
phenomenological model depend only weakly on a T-dependent DOS, as long as
the peak width in the DOS does not vary faster than temperature. Further, it
would be interesting to investigate e-ph thermalization in Kondo insulators.
Namely, due to the presence of the gap near E$_{F}$ one would expect effects
similar to the Rothwarf-Taylor bottleneck observed in superconductors\cite%
{Kabanov}. Secondly, even more interesting effects are expected due to the
strong reduction of the screening at low frequencies (below the gap) which
could lead to non-adiabatic phonons.

In conclusion, we have utilized ultrafast optical spectroscopy to study the
dynamics of photoexcited quasiparticles in HF compounds. We have observed a
divergence in the e-ph thermalization time at low temperatures. We argue
that the dramatic hundred-fold increase in the relaxation time at low
temperatures in Yb\textit{X}Cu$_{4}$ (and the lack of this quasi-divergence
in the non-HF Lu\textit{X}Cu$_{4}$ analogs) results from the largely
increased DOS at $E_{f}$ coupled with strongly suppressed scattering of
heavy-electrons by phonons.

We thank Kaden Hazzard for the specific heat data. This work was supported
by US DOE.

\subsection{Figure Captions}

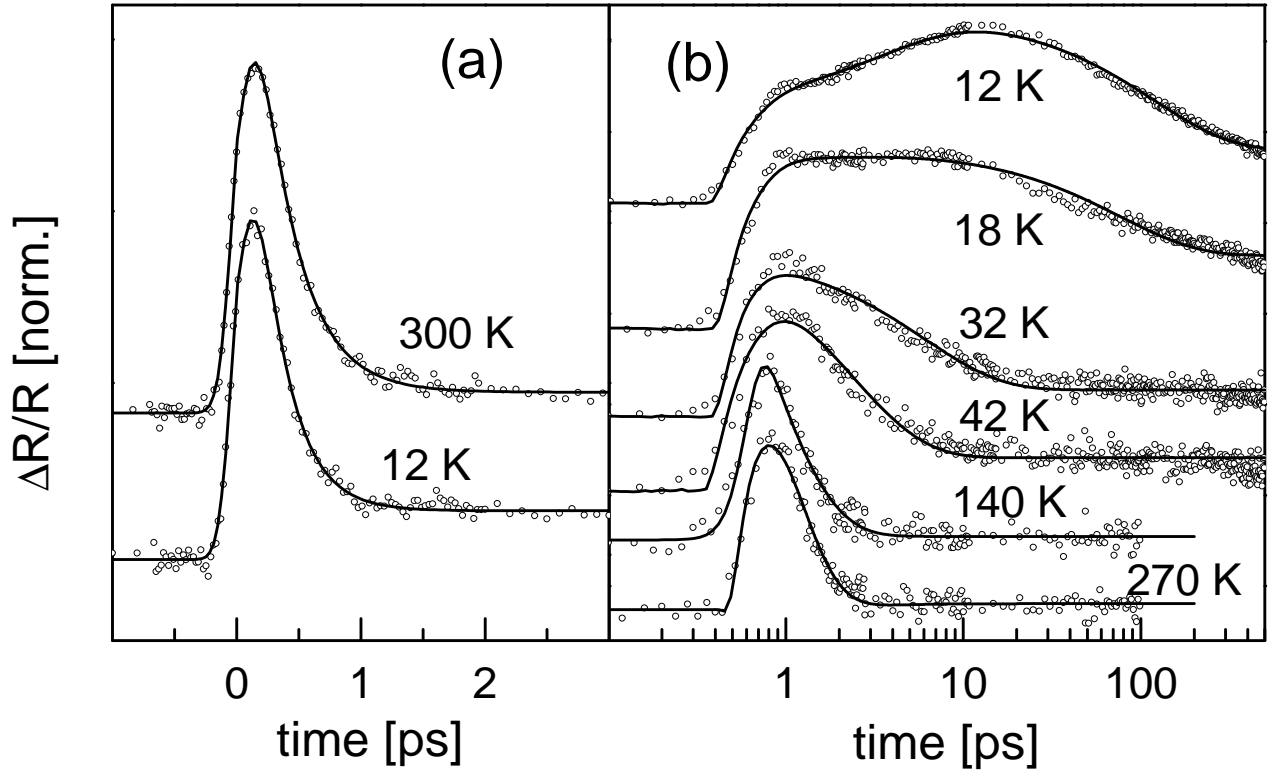
\textbf{Figure 1}

Normalized photoinduced reflectivity data (solid symbols) on (a)\ LuAgCu$%
_{4} $ and (b) YbAgCu$_{4}$ at various temperatures, together with best fits
(see text) to the data - solid lines. The data have been vertically shifted
for clarity.

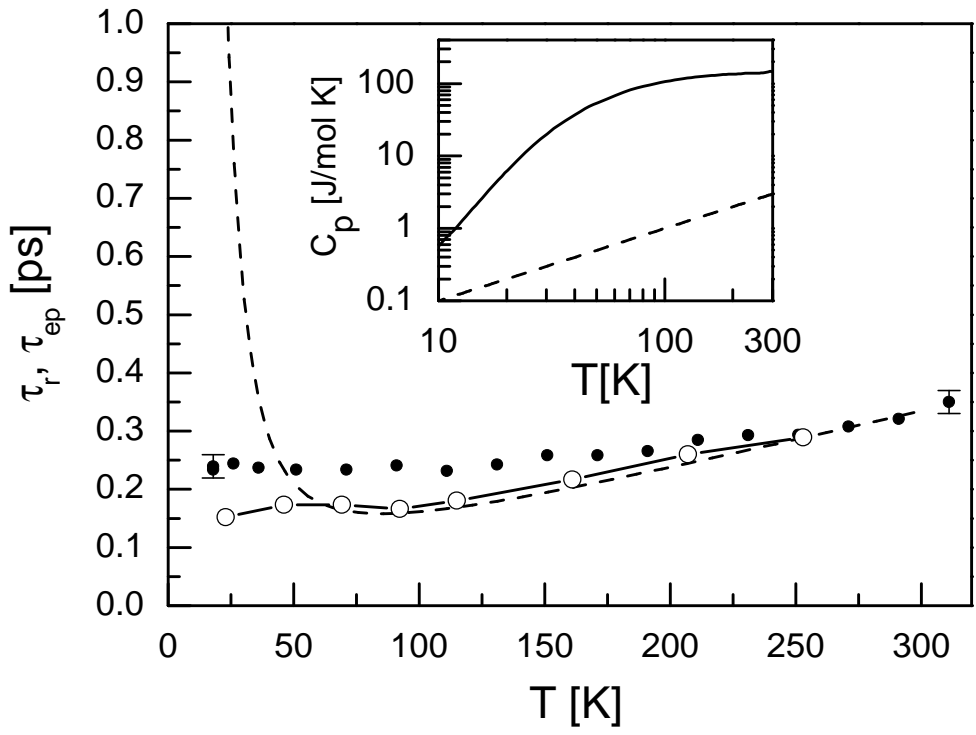
\textbf{Figure 2}

a) Temperature dependence of relaxation time $\tau _{r}$ on LuAgCu$_{4}$
(solid circles), together with the TTM prediction (dashed line) and the
result of the numerical simulation (open circles) including the non-thermal
electron distribution. Inset: $C_{e}$ (dashed) and $C_{l}$ (solid line) of
LuAgCu$_{4}$.

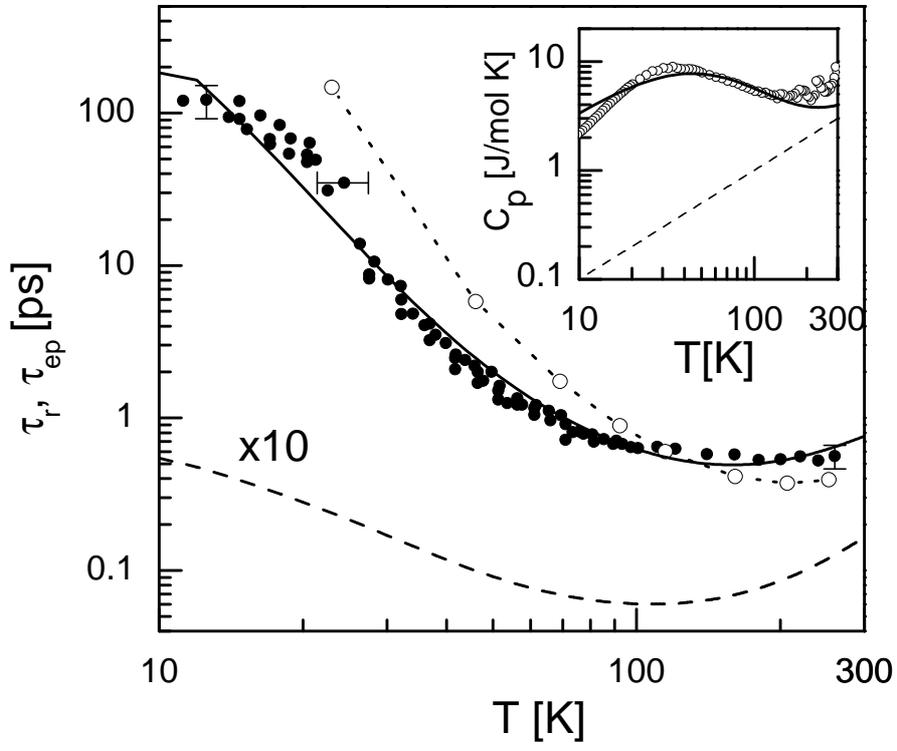
\textbf{Figure 3}

T-dependence of $\tau _{r}$ on YbAgCu$_{4}$ (solid circles), together with
the corresponding $\tau _{ep}$ (multiplied by 10 for the presentation
purpose) from simple TTM prediction (dashed line). Assuming suppressed
scattering of heavy electrons by phonons due to $v_{F}<v_{s}$ condition,
very good agreement with the data is obtained: open circles present the
result of numerical simulation, while solid line presents the TTM simulation
- see text. Inset: experimentally determined $C_{e}(T)$ for YbAgCu$_{4}$
(open circles), together with calculated $C_{e}(T)$ based on the model $%
D_{e}(\epsilon )$ - solid line. $C_{e}(T)$ of LuAgCu$_{4}$ (dashed line) is
added for comparison.


\begin{thebibliography}{99}
\bibitem{note2} The anomalous rise-time dynamics may reflect some kind of a
bottleneck in the e-e thermalization due to complicated low-energy
electronic structure, e.g. hybridization gap.
\bibitem{numerics1} The simulation shows that $\tau _{ep}$ does not depend
on the details of the initial condition\cite{Groenevald,fullana}.
\bibitem{numerics2} While $K_{ee}$ is not uniquely determined, fitting $\tau
_{ep}$ to the experimental data requires $K_{ee}/K_{ep}<7\times 10^{3}$. If $%
K_{ee}/K_{ep}\geq 7\times 10^{3}$ the e-e scattering rate is fast enough for
the electron system to quickly reach thermal distribution, resulting in $%
\tau _{ep}\propto T^{-3}$ behavior predicted by the TTM. Indeed, the
analysis of the time evolution of $df/dt$ shows that if $K_{ee}/K_{ep}=700$,
then $\tau _{ee}>\tau _{ep}$ at low temperatures\cite{fullana}.
\bibitem{note3} The absolute values of the e-ph coupling in both numerical
and the TTM simulations were smaller than the corresponding value for LuAgCu$%
_{4}$ by about half in terms of the scattering matrix element.
\end{thebibliography}
\end{document}